\documentclass[conference]{IEEEtran}

\usepackage{hyperref}
\hypersetup{draft}

\usepackage{hyperref}
\usepackage[font=small,skip=10pt]{caption}
\usepackage{amsmath}    
\usepackage{algorithm}
\usepackage{algorithmic}
\usepackage{multirow}
\usepackage{svg}
\usepackage{multicol}
\usepackage{caption}
\usepackage{float}
\usepackage{chemformula}
\usepackage{color}

\usepackage{tikz}
\tikzset{
main node/.style={inner sep=0,outer sep=0},
label node/.style={inner sep=0,outer ysep=.2em,outer xsep=.4em,font=\scriptsize,overlay},
strike out/.style={shorten <=-.2em,shorten >=-.5em,overlay}
}

\usepackage{graphicx}
\DeclareGraphicsExtensions{.pdf,.jpeg,.png,.fig, .emf}
\usepackage{subfigure}
\usepackage{epstopdf}
\usepackage{epsfig}
\usepackage{subfigure}
\usepackage{epstopdf}
\usepackage{epsfig}
\usepackage{cancel}
\usepackage{amsmath}

\usepackage{RobStd}
\HideNotes
\usepackage{makecell}
\usepackage{booktabs}
\usepackage{multirow}
\usepackage{tabularx}

\newcommand\eat[1]{}
\newcommand{\name}{{CAMASim}\xspace}

\usepackage{wasysym}

\begin{document}


\title{CAMASim: A Comprehensive Simulation Framework 
for Content-Addressable Memory \\based Accelerators}



\author{
\IEEEauthorblockN{Mengyuan Li, Shiyi Liu, Mohammad Mehdi Sharifi, X. Sharon Hu}
\IEEEauthorblockA{Department of Computer Science and Engineering\\
University of Notre Dame, Notre Dame, IN, USA, 46556\\
Emails: \{mli22, sliu33, msharifi1, shu\}@nd.edu}}



\maketitle

\begin{abstract}
Content addressable memory (CAM) stands out as an efficient hardware solution for memory-intensive search operations by supporting parallel computation in memory. However, developing a CAM-based accelerator architecture that achieves acceptable accuracy, while minimizing hardware cost and catering to both exact and approximate search, still presents a significant challenge especially when considering a broader spectrum of applications. This complexity stems from CAM's rapid evolution across multiple levels—algorithms, architectures, circuits, and underlying devices. This paper introduces \name, a first comprehensive CAM accelerator simulation framework, emphasizing modularity, flexibility, and generality. \name establishes the detailed design space for CAM-based accelerators, incorporates automated functional simulation for accuracy, and enables hardware performance prediction, by leveraging a circuit-level CAM modeling tool. This work streamlines the design space exploration for CAM-based accelerator, aiding researchers in developing effective CAM-based accelerators for various search-intensive applications.
\end{abstract}

\section{Introduction}

Mitigating the memory wall challenge inherent in the Von-Neumann architecture has been a long-standing architecture research focus. In recent years, the rapid development of machine learning and artificial intelligence algorithms, which is ever more memory-intensive, has highlighted the demand for investigating in-memory computing (IMC) architectures. Researchers have introduced various IMC kernels to deal with different computation and memory patterns. Among these IMC kernels, content addressable memory (CAM), also referred to as associative memory, has gained special attention. Different from IMC kernels like crossbars targeting matrix-vector multiplications, CAM emerges as a potential solution for applications requiring efficient search functionalities, catering to both exact and approximate matches. Notably, CAM-based application-specific accelerators have been designed to enhance a myriad of applications, e.g., computer vision~\cite{karunaratne2021robust}, recommendation system~\cite{imars}, DNA sequencing~\cite{seedvote}, and reinforcement learning~\cite{li2022associative} showcasing significant performance advantages over conventional computing platforms. 


However, designing efficient CAM accelerators for diverse applications poses significant challenges, primarily due to the large and complex design space of CAM-based accelerators. From bottom-up, at the memory device level, both CMOS and non-volatile memory (NVM) devices can be used to construct CAM cells, each has its pros and cons. At the circuit level, not only different device choices may lead to different circuit design options but also diverse sensing circuits may be needed/selected based on different match types (e.g., exact match or best match) and precision/performance requirements. Furthermore, at the architectural level, the sizes of the basic CAM array and the peripherals used to merge array results can have significant impact on both application-level accuracy and area/latency/energy. Last but not least, application choices directly influence the choices of architectures as well as match types. The decisions at all these levels are interdependent and eventually determine the final application accuracy and the system hardware performance including latency, energy and area etc.

To design a superior in-memory search accelerator for a given application (or a set of applications), it is imperative to efficiently explore the intricate design space of CAM-based accelerators. Towards this end, a comprehensive CAM evaluation framework which can accurately predict application-level accuracy as well as hardware performance is highly desirable. Though several CAM evaluation frameworks, such as NVSim-CAM~\cite{nvsim} and EvaCAM~\cite{evacam}, have been proposed. They only focus on circuit- and device-level modeling to derive hardware performance, thus ignoring the impact of application- and architecture-level decisions. More importantly, these tools do not evaluate application accuracy, which may be equally or even more important than hardware performance. 


To fill this void, in this work, we introduce a comprehensive simulation framework, referred to as \name~\footnote{The source code of \name version 1.0 is publicly available at \href{https://github.com/menggg22/CAMASim}{https://github.com/menggg22/CAMASim}.
}, to efficiently model and evaluate CAM-based in-memory search accelerator designs. \name considers the design choices at different levels, including architectural considerations such partition and merge schemes, circuit parameters like the sensing circuit limit, diverse CAM cell types and device variations, etc. \name provides an easy-to-use Python interface for search-intensive algorithms at the application level, facilitating seamless integration and testing. The design of ~\name focuses on modularity, flexibility, and generality. 

To our best knowledge, \name is the first CAM evaluation framework that considers both accuracy and hardware performance across multiple design layers. The main contributions of this work include: (i) introducing an automated functional simulation flow for accuracy evaluation with explicit consideration of hardware non-idealities; (ii) enabling hardware performance simulation, compatible with low-level circuit modeling tools and accommodating various peripherals; (iii) supporting multi-level configurations, accommodating diverse applications, architectures, circuit and underlying devices.

\section{Background}\label{sec:background}
Below we introduce CAM basics and then review existing efforts on the simulation frameworks for CAM-based accelerators.

\begin{table}[t]
\renewcommand{\arraystretch}{1.5}
\centering
\begin{tabularx}{0.465\textwidth}{>{\centering\arraybackslash}m{0.10\linewidth}|>{\centering\arraybackslash}m{0.08\linewidth}|>{\centering\arraybackslash}m{0.12\linewidth}|>{\centering\arraybackslash}m{0.12\linewidth}|>{\centering\arraybackslash}m{0.12\linewidth}|>{\centering\arraybackslash}m{0.14\linewidth}}

\toprule
~ & \textbf{Dim}  & \textbf{\#Entries} & \textbf{Distance} & \textbf{Match Type} & \textbf{Data Type}  \\
\hline

\hline
\textbf{\shortstack{DRL\\\cite{li2022associative}}} & 32 & 2-10k   & Hamm. & Exact & 1 bit\\
\hline
\textbf{\shortstack{HDC\\\cite{kazemi2022achieving}}}  & 75 - 784 & 2-26 & L2 & Best & 1 or more bits \\
\hline
\textbf{\shortstack{MANN\\\cite{laguna2023fewshot}}} & 64-128 & 10-100 & L2 & Best & 1 or more bits\\
\hline
\textbf{\shortstack{DNA\\\cite{seedvote}}} & 64 & 10k+ & Hamm. & Exact/Best  & 1 bit\\
\bottomrule
\end{tabularx}
\caption{Application-level design in several reported CAM-based accelerators. (Hamm.: Hamming)}
\label{table:application}
\end{table}

\label{sec:related}
\subsection{Content-addressable Memory}
CAM facilitates fast and energy-efficient searches within memory without the need of moving data to the processing unit. The core operations of CAM include search and write. During search, a CAM array simultaneously identifies the entry matching the query, while during writing, it stores data in the corresponding CAM entries. By executing parallel searches for a query across all data stored in memory, CAM accomplishes search in constant time (O(1)).  


CAM can be built with different device technologies, including both conventional CMOS and emerging NVM devices. The conventional 16T CMOS-based CAM incurs area and leakage penalties~\cite{ni2019ferroelectric}. Emerging NVM technologies such as ReRAM and FeFET have led to several CAM designs that offer low-power, high-speed, and high-density benefits~\cite{ni2019ferroelectric, skytcam}. 

Depending on the data representation in the CAM cell, CAM can be categorized binary CAM (BCAM), storing `1' or `0'; ternary CAM (TCAM), capable of storing `1', `0', and a `don't care' state; multi-bit CAM (MCAM), accommodating multiple-bit data in a cell; and analog CAM (ACAM), designed to store analog ranges. A CAM array can incorporate a range of sensing circuits to facilitate different match types, including (i) exact match, which identifies the row(s) with every cell matching the query, (ii) best match, which
determines the row more similar to the query, (iii) threshold match, which identifies rows that the distance to the query falls below a threshold. We refer readers to~\cite{hu2021memory} for more details about CAM.

Prior work has introduced CAM-based in-memory search accelerators based on different devices, cell and array designs for applications in various domains, including Deep Reinforcement Learning (DRL)~\cite{li2022associative}, hyperdimensional computing (HDC)~\cite{kazemi2022achieving}, memory-augmentation neural networks (MANN)~\cite{laguna2023fewshot}, DNA Sequencing~\cite{seedvote}. Table~\ref{table:application} lists the corresponding CAM-based accelerators with the application-level considerations such as distance function, and match type and data type. Later in Sec.~\ref{sec:ds}, we will establish a comprehensive design space for CAM-based accelerators.



\begin{table}[t]
\footnotesize
\renewcommand{\arraystretch}{1.5}
\centering
\begin{tabularx}{\linewidth}{>{\centering\arraybackslash}m{0.23\linewidth}|>{\centering\arraybackslash}m{0.14\linewidth}|>{\centering\arraybackslash}m{0.14\linewidth}|>{\centering\arraybackslash}m{0.11\linewidth}|>{\centering\arraybackslash}m{0.15\linewidth}}
\toprule
~ & \textbf{NVSim-CAM}~\cite{nvsim} & {\shortstack{\textbf{EvaCAM} \\ ~\cite{evacam}}} & {\shortstack{\textbf{XTime} \\ ~\cite{pedretti2023x}}} & \textbf{\name (ours)} \\
\hline
Accu. Sim. & \Circle & \Circle & \CIRCLE & \CIRCLE\\
\hline 
Perf. Sim.  & \CIRCLE & \CIRCLE & \CIRCLE & \CIRCLE \\
\hline
App. Config. & \Circle & \Circle & \LEFTcircle & \CIRCLE\\
\hline
Arch. Config.  & \Circle & \LEFTcircle & \LEFTcircle & \CIRCLE \\
\hline
Circ./Dev. Config. & \LEFTcircle & \CIRCLE & \Circle & \CIRCLE\\
\bottomrule
\end{tabularx}
\caption{Existing CAM evaluation frameworks comparison. \Circle: no coverage, \LEFTcircle: some support with limited generality, \CIRCLE: comprehensive consideration.}
\label{table:related}
\end{table}

\subsection{Existing CAM Simulators}
There are several prior efforts in CAM simulation framework, as shown in Table~\ref{table:related}. To understand the landscape of CAM evaluation advancements, we assess these tools based on their capabilities, encompassing accuracy simulation (accu. sim.) and hardware performance simulation (perf. sim.), as well as configurability (config.) across application (app.) /architecture (arch.) /circuit (cric.) and device (dev.) levels.

Early effort like NVSim-CAM~\cite{nvsim} focuses on NVM-based TCAMs and has limited device and circuit-type support. A recent work, EvaCAM~\cite{evacam}, extends the circuit-level support to more NVM devices like FeFETs and models more cell types including ACAM and MCAM. But these tools are confined to hardware performance. XTime~\cite{pedretti2023x} develops the evaluation tool with both application-level accuracy and hardware performance but only focused on the random forest task and the corresponding ACAM-based architecture, which cannot be applied to other application scenarios. Thus, a comprehensive simulator for both application accuracy and hardware performance is highly needed to advance the continued development of CAM-based accelerators.


\section{\name}
\label{sec:tool}
In this section, we present our \name framework, beginning with a high-level overview of the tool's purpose and functionality. Subsequently, we delve into the supported design space, elucidating \name's configurability on each level. Following this, we provide an in-depth introduction of two main components: the functional simulator and the performance evaluator.

\begin{figure*}
    \centering
    \includegraphics[width=0.99\textwidth]{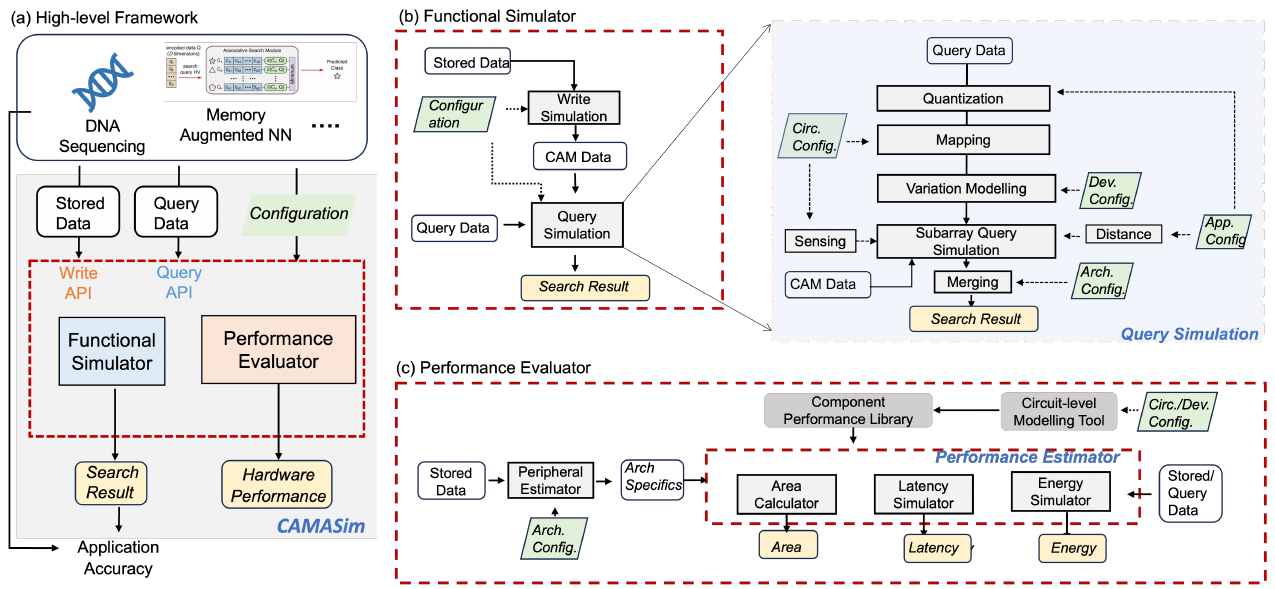}
     \caption{\name framework. (a) High-level framework. (b) Functional simulator. (c) Performance evaluator.}
    \label{fig:framework}
\end{figure*}

\begin{table}[thb]
\footnotesize
\renewcommand{\arraystretch}{1.5}
\centering
\begin{tabularx}{\linewidth}{>{\centering\arraybackslash}p{0.15\linewidth}|>{\centering\arraybackslash}p{0.4\linewidth}|>{\centering\arraybackslash}X}
\toprule
~ & \textbf{Parameter} & \textbf{Value} \\
\hline
\multirow{4}{*}{\shortstack{\textbf{App.} \\ \textbf{Config.}}} & Distance Function & Hamm./L1/L2 \\
\cline{2-3}
& Match Type & Exact/Best/Threshold \\
\cline{2-3}
& Match Parameter & Int \\
\cline{2-3}
& Data Type & Int \\
\hline
\multirow{4}{*}{\shortstack{\textbf{Arch.} \\ \textbf{Config.}}} & Subarrays per Array & Int \\
\cline{2-3}
& Arrays per Mat & Int \\
\cline{2-3}
& Mats per Bank & Int \\
\cline{2-3}
& Horizontal Merge Type & Voting/AND\\
\cline{2-3}
& Vertical Merge Type & Comparator/Gather \\
\hline
\multirow{4}{*}{\shortstack{\textbf{Circ.} \\ \textbf{Config.}}} & Row & Int \\
\cline{2-3}
& Column & Int \\
\cline{2-3}
& Cell Type & B/T/A/MCAM \\
\cline{2-3}
& Sensing Circuit Type & Exact/Best/Threshold \\
\cline{2-3}
& Sensing Limit & FP32 \\
\hline
\multirow{3}{*}{\shortstack{\textbf{Dev.} \\ \textbf{Config.}}} & Device Type & String \\
\cline{2-3}
& Variation Type & D2D/C2C \\
\cline{2-3}
& Variation Specification & Stat./Exper. \\
\cline{2-3}
& Variation STD & FP32 \\
\bottomrule
\end{tabularx}
\caption{\name configuration parameters.}
\label{table:config}
\end{table}

\subsection{Overview of \name}
A high-level depiction of \name is shown in Fig.~\ref{fig:framework}(a). Given a specific search-intensive application, \name generates the search results from the CAM-based accelerator for predicting application accuracy, and the accelerator performance. The input to \name comprises three main elements: \textit{stored data} (data to be stored in CAM), \textit{query data} (query input), and a detailed \textit{configuration} file (describing the accelerator's design choices at multiple levels). The user defines stored data and query data, provided to \name via its dedicated write and query APIs. The comprehensive \textit{configuration} file detailing the design choices at each level configures \name for different simulation setups. The functional simulator within \name supports automated in-memory search simulation, generating the search results, i.e., the match entry indices, for exact, best and threshold matches. The performance evaluator, building on top of the underlying circuit-level modelling tool, generates predicted performance values encompassing latency, energy and area information. 

\subsection{CAM-based Accelerator Design Space}\label{sec:ds}
\name offers multi-level configurations for CAM-based accelerator designs, enabling users to explore the design space via the \textit{configuration} file. Here we describe the design space considered in \name as specified by the configurable parameters across the application, architecture, circuit, and device levels shown in Table~\ref{table:config}.

\textbf{Application level:} As can be seen from Table~\ref{table:application}, diverse applications have distinct requirements, encompassing considerations such as distance function, match type and data type. To accommodate a wide range of applications, \name offers adjustable parameters using the application configuration (see the rows corresponding to \textit{app. config.} in Table~\ref{table:config}). This includes provisions for specifying the distance function, match type, and match parameters (number of neighbors or threshold value) and data type (number of bit). 

\begin{figure}[t]
    \centering
    \includegraphics[width=\linewidth]{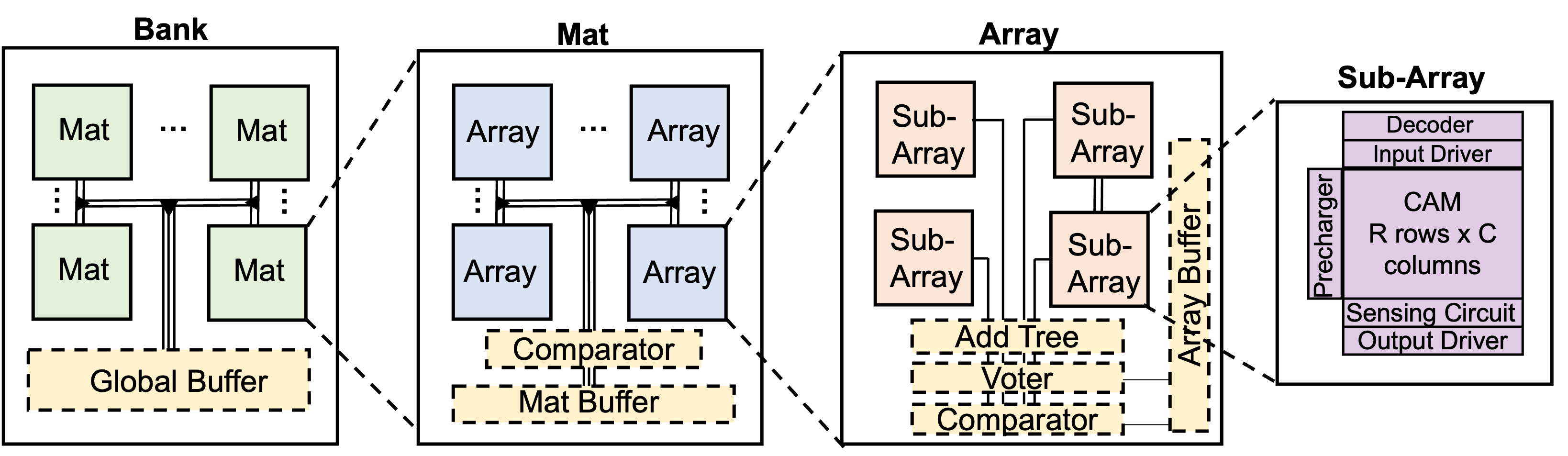}
     \caption{Hierarchical structure of CAM-based accelerator design.}
    \label{fig:arch}
    \vspace*{-5mm}

\end{figure}

\textbf{Architecture level:} To capture the architecture-level design choices, we adopt a general and flexible architecture design for CAM-based accelerators, visualized in Fig.~\ref{fig:arch}. The basic computational unit is a CAM subarray with $R$ rows and $C$ columns. The architecture uses a typical hierarchical four-layer structure, bank-mat-array-subarray, to deal with potentially large stored data. Each layer consists of multiple lower-layer blocks which operate in parallel and the peripherals for merging the search results, allowing the comparison of query data to all stored data simultaneously. The user can configure the number of blocks at each layer within the architecture configuration as shown in \textit{arch. config.} rows in Table~\ref{table:config} including subarrays per array, arrays per mat and mats per bank.

\begin{figure}
    \centering
    \includegraphics[width=\linewidth]{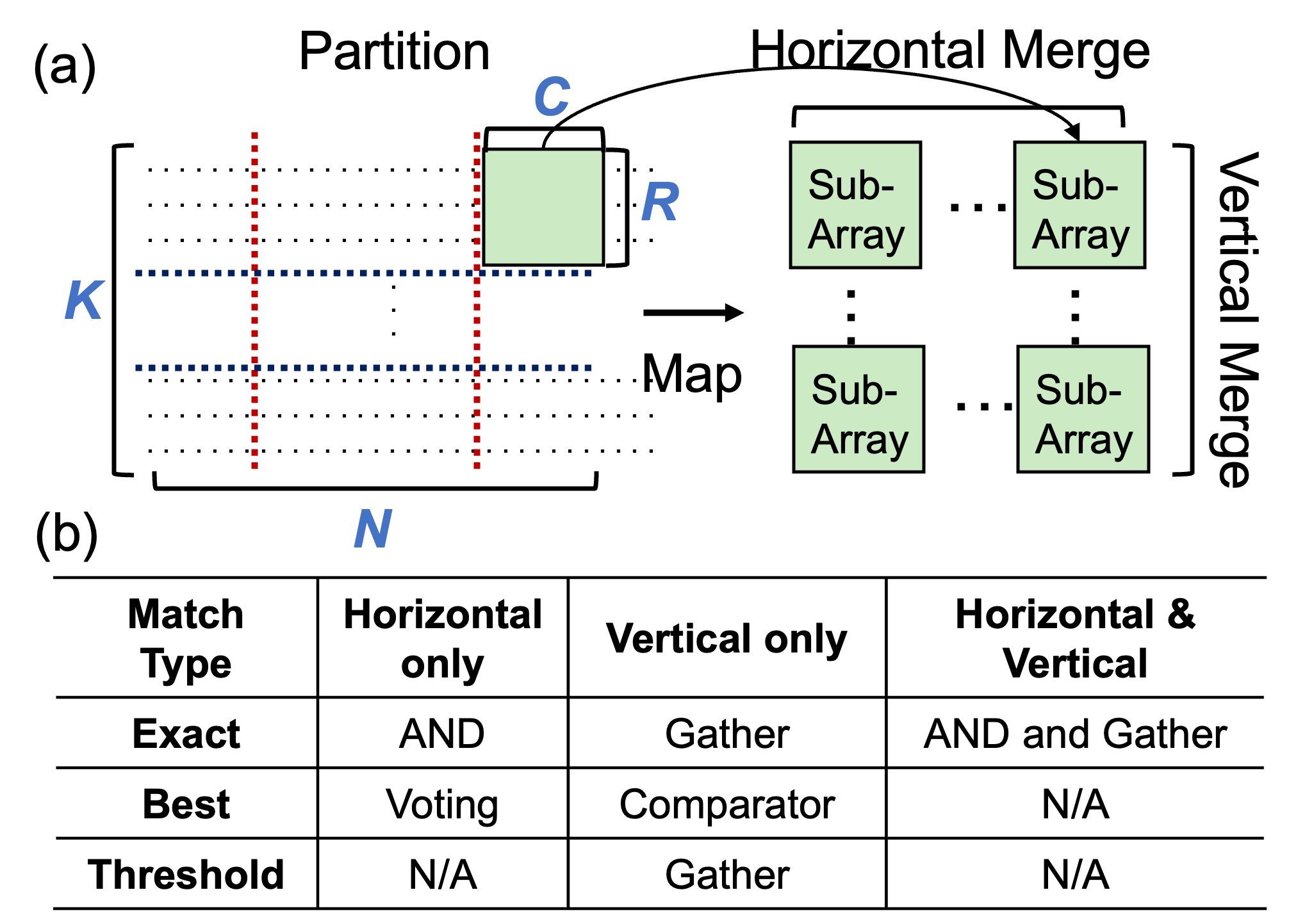}
    \caption{ (a) Illustration of the partition and merge problem in CAM-based accelerator. (b) Existing horizontal and vertical merge schemes for exact/best/threshold matches.}
    \label{fig:partition}
        \vspace*{-5mm}

\end{figure}

Efficiently obtaining the final search results requires well-designed peripheral circuits for merging search results from the multiple low-layer blocks. Herein lies a critical architectural consideration in CAM-based accelerator design, the \textbf{partition and merge} problem, shown in Fig.~\ref{fig:partition}(a). For the stored data characterized by $N$ dimensions and $K$ entries, given subarrays of $R$ rows and $C$ columns we partition and map the stored data into multiple subarrays. After querying on each subarray in parallel, a scheme is needed to to merge the subarray search results into the application-level search results while ensuring accuracy and low hardware cost. 

Depending on the relationship between the stored data size and the CAM subarray size, merge problems can be broadly categorized into two types: horizontal merge (when $N>C$) and vertical merge (when $K>R$). In Fig.~\ref{fig:partition}(b), we summarize existing merge schemes developed for various application-level match types. In horizontal merge, each subarray only contains a portion of the long vector. Consequently, horizontal merge might introduce errors in search results depending the match type. For the exact match, an AND operation over all subarray results can yield the exact match across the entire vector. However, in the case of best match, the search result from each subarray represents the most similar entry over a portion of the vector. Previous work~\cite{kazemi2022achieving} introduced the voting scheme to identify the approximate best match across the entire vector. It is worth noting that there's no existing efficient solution to deal with the horizontal merge problem for threshold match. 
The vertical merge problem involves combining results from multiple entries. For both exact and threshold matches, the vertical merge scheme of gathering all search results from all subarrays simply produces the correct application-level search results. However, for best match, the comparator-based vertical merge is required to compare subarray results and obtain the best match among all stored data. Moreover, currently there is no CAM-based accelerator design that addresses both horizontal and vertical merge problems simultaneously for best and threshold matches.

For each merge scheme, peripheral circuits need to be designed at each layer. As a example, the yellow shaded blocks in Fig.\ref{fig:arch} illustrates the peripherals designed for the voting-type horizontal merge scheme, proposed for HDC~\cite{kazemi2022achieving}. Currently, \name supports all the existing merge schemes detailed in Fig.~\ref{fig:partition}(b) (see the last two rows in the~\textit{arch. config.}), which enables users to explore and assess the impact of horizontal and vertical merge schemes on the application-level accuracy and hardware performance. It is worth noting that more efficient merge schemes are still to be developed for both horizontal and vertical merge problems when it comes to best match and threshold match. \name can readily accommodates user-defined new merge schemes, ensuring that \name's continued value in the evolvement of CAM-based accelerator designs.

\textbf{Circuit level:} At the circuit level, we focus on subarray circuit design exploration with two main design considerations (see the \textit{circ. config.} section of Table~\ref{table:config}). The first includes the size of subarray, i.e., the row and column configuration, and the CAM cell type. The second is the CAM subarray's match type, which is determined by the design of the sensing circuit. Also, the sensing limit (SL) of the sensing circuit is considered here, which determines the minimum voltage/current difference can be detected by the sensing circuit.

\textbf{Device level:} As we discussed in Sec.~\ref{sec:background}, CAM can be designed with different devices, resulting in varying hardware performance. \name offers the choices of device types in \textit{dev. config.} as in Table~\ref{table:config}. Additionally, to facilitate device variation aware simulations, \name considers two types of variations: device-to-device (D2D) variation and cycle-to-cycle (C2C) variation. Two types of variation specification are included: (i) statistical (stat.) Gaussian variation, offering tunable standard deviation (STD); (ii) experimental (exper.) variation with variation distributions measured from fabricated chips. The choices of device variation types and specifications are included in~\textit{dev. config.} in Table~\ref{table:config} to facilitate exploration and analysis of different device variation options.

\subsection{Functional Simulator}
To predict application-level accuracy, \name generates the \textit{search result}—the indices of match entries across all stored data for a given query data. The result depends upon multiple factors spanning from the application to the device levels. Any alterations in design choices can influence the final outcome, underscoring the necessity for a functional simulator that supports modular processing for isolation of concerns.

Fig.~\ref{fig:framework}(b) illustrates the functional simulation flow. Initially, the \textit{stored data} undergoes processing via the Write Simulation module, generating \textit{CAM Data} distributed into the subarrays. Subsequently, the {Query Simulation} module receives the \textit{query data} and produces the search results based on \textit{CAM Data}. For both query and stored data, several key submodules come into play: (i) Quantization: Processes data to fit the underlying cell data type. (ii) Mapping: Maps data to corresponding subarrays. (iii) Variation Modeling: Incorporating device variation to the stored data. (iv) Subarray Query Simulation: Executes search on each subarray. (v) Merging: Aggregates subarray search results to generate the final search results. Note that the Write Simulation shares the quantization, mapping and variation modelling submodules with the Query Simulation. Due to the page limit, this part is omitted in Fig.~\ref{fig:framework}. Next we delve into detailed discussions of these submodules.

The \textbf{Quantization} submodule processes both stored data and query data from the application level to match the requisite data type representations specified by the user in \textit{app. config.} to specifically addressing constraints such as the limited number of bits in TCAM and MCAM. This module employs the linear quantization technique to transform the data into either binary or 2/3-bit representations. Other quantization techniques can be readily added as choices in the submodule.

\begin{table*}[h]
\centering
\renewcommand{\arraystretch}{1.5}

\begin{tabularx}{\textwidth}{@{}>{\centering\arraybackslash}p{1.8cm}|*{2}{>{\centering\arraybackslash}X|}>{\centering\arraybackslash}p{1.5cm}|*{2}{>{\centering\arraybackslash}X|}>{\centering\arraybackslash}p{1.12cm}|*{5}{>{\centering\arraybackslash}X|}>{\centering\arraybackslash}X}



\toprule
& \textbf{Match Type} & \textbf{Dist.} & \textbf{Cell Type}  & \textbf{Device}& \textbf{\# Subarray} & \textbf{Subarray Size} & \textbf{Acc.  (pub.)} & \textbf{Acc. (sim.)} & \textbf{Latency (pub.)}  & \textbf{Latency (sim.)} & \textbf{Energy (pub.)} & \textbf{Energy (sim.)} \\
\hline

\textbf{DRL~\cite{li2022associative}} & Exact & Hamm. & TCAM & CMOS & 64 & 64x64 & 173.25 & 169.50 & 1.0us & 0.95us& /& 46.0uJ\\
\hline

\textbf{MANN~\cite{laguna2023fewshot}} & Best & L2 & MCAM-3b & FeFET & 8 & 32x64 & 94.5\%& 95.0\%& 6.5ns & 6.4ns & 16.6pJ & 17.7pJ\\
\hline
\textbf{HDC~\cite{kazemi2022achieving}} & Best & L2 & MCAM-2b & FeFET  & 16 & 32x128 & 94.6\% & 95.1\% & 12.2ns & 12.8ns& 269.0pJ & 252.0pJ\\
\bottomrule

\end{tabularx}
\caption{Validated accuracy, latency, energy of \name results (sim.) against the reported data in published work (pub.) of CAM-based accelerator on HDC, MANN, DRL task.}
\label{table:accuracy}
\vspace*{-5mm}
\end{table*}

The \textbf{Mapping} and \textbf{Merging} submodules deal with the challenge of partition and merge problem in the CAM-based accelerator. The mapping submodule first partitions stored data into multiple subarrays according to the subarray size, and segments the query data according to the number of columns in each subarray. Then it maps these subarrays onto a 2-D grid for subsequent horizontal and vertical merges. Within the merging submodule, results from each subarray are combined following the predefined merge schemes specified by the \textit{arch.} configuration. Currently, the submodule incorporates the merge schemes listed in Table.~\ref{fig:partition}(b) such as horizontal voting scheme for best match. Both the mapping and merging submodules are extensible to accommodate future schemes, fostering a continuous pursuit aiming at improving accuracy for CAM-based accelerators.

The \textbf{Variation Modelling} submodule models the impact of device variations on the stored data. The variation type and specification are given in the \textit{dev.} configuration. The submodule incorporates the D2D variation through a one-time addition, and C2C variation via dynamic addition for each query. The configurability of variation types as well as how variation is specified empowers users to assess the impact of diverse device variation on the end-to-end accuracy, facilitating a comprehensive evaluation of system behavior for the given application.


The \textbf{Subarray Query Simulation} submodule simulates subarray-level search operations to obtain the search results for each subarray. This submodule uses the subarray size and sensing limit from the \textit{circ.} configuration. SL, determined by the sensing circuit, is defined as the smallest voltage/current difference detectable by the SA. In \name, SL is treated as a configurable parameter and the submodule outputs all entries in each subarray within SL. E.g., for best match type search, the second closest entry within SL is also detected as a match.

\subsection{Performance Evaluator}

Fig.~\ref{fig:framework}(c) illustrates the workflow of the performance evaluator responsible for generating the CAM-based accelerator performance such as latency, energy and area based on the provided input. To facilitate various architectural design choices, the evaluation process comprises two key stages: (i) architecture specifics estimation and (ii) performance prediction.  

First, \name estimates the architecture specifics including the number of computing block and the peripheral circuit type and size at each layer for a CAM architecture as depicted in Fig.~\ref{fig:arch}. Currently, \name assumes all stored data fitting in the CAM and determines the number of computing block at the bank, mat, and array layers based on \textit{arch. config.} details and the size of stored data. Depending on the merge scheme, the tool estimates peripherals' circuit type and size through a \textit{peripheral estimator}. An instance of this can be seen in Fig.~\ref{fig:arch}. For this merge scheme, when provided with the number of arrays per mat, the \textit{peripheral estimator} performs estimations for the requisite number of comparators and determines buffer size accordingly.

For performance prediction, by leveraging the CAM architecture specifics, \name empolys a performance estimator (see Fig.~\ref{fig:framework}) to predict the CAM performance corresponding to the write and search operation. The process follows a hierarchical approach: bank-mat-array-subarray. At each layer, the performance of CAM, peripherals, and interconnects are estimated. At the subarray level, \name integrates either external circuit-level CAM modeling tools (e.g., EvaCAM~\cite{evacam}) or actual SPICE simulation results to generate the performance values, ensuring compatibility with diverse underlying CAM cell designs based on different devices. Moreover, a catalog of commonly-used peripheral designs are included in \name to enable users to select peripherals for their merge schemes. More implementation details will be covered in Sec.~\ref{sec:evaluation}. The performance estimator evaluation yields the detailed hardware performance for the given CAM accelerator, enhancing understanding and optimization potential at the architecture level.

\section{Evaluation}
\label{sec:evaluation}
\subsection{Validation}

We first validate \name against several reported CAM-based accelerator designs by comparing the \name generated latency, energy, accuracy with the results reported in previous work. Note that due to the unavailability of area information in the literature, we do not include area validation here. We configure \name for applications including MANN~\cite{laguna2023fewshot}, HDC~\cite{kazemi2022achieving}, DRL~\cite{li2022associative}. To ensure the correctness of the reproduction, we adopt the same application, architecture, circuit and device setups reported in the respective literature. 

As in the previous work, for hardware performance, \name employs the underlying CAM and peripheral circuits built with 22nm technology. The CAM components operate at a maximum clock rate of 150MHz. The CAM performance is retrieved from the circuit modelling tool EvaCAM~\cite{evacam}. For the CAM and sensing circuit designs, e.g., FeFET MCAM with best match sensing circuit, not supported by EvaCAM, we include the circuit performance obtained from SPICE simulation. The performance values of peripheral circuits, such as comparators, adders, and registers, have been derived from pre-RTL simulations conducted in ALADDIN \cite{shao2014aladdin}. The performance evaluation of interconnects are based on RC estimations as implemented in NVSIM~\cite{dong2012nvsim}. 

The simulation results from \name, along with the results reported in prior work, are summarized in Table~\ref{table:accuracy}. In the case of DRL, the accuracy  (i.e. test score) in simulation exhibits a small deviation, attributed to the randomness inherent in the implemented sampling operation by the CAM. In contrast, MANN demonstrates a slightly higher accuracy (95.0\% versus 94.5\%). This difference is attributed to \name's consideration of subarray search operations, which were overlooked in the previous implementation. Additionally, HDC exhibits accuracy with a minor deviation, stemming from slight disparities in the simulation environment versions.

Regarding the hardware performance across the three tasks, \name reports latency with a 1.5\% to 5\% deviation and energy with approximately a 6\% deviation. These deviations primarily arise from minor differences in the underlying circuit performance. Nevertheless, \name efficiently reports application-level accuracy and performance, aligning closely with previously reported data.


\begin{figure}[b]
    \centering
        \vspace*{-5mm}
\subfigure{\includegraphics[width=0.49\linewidth]{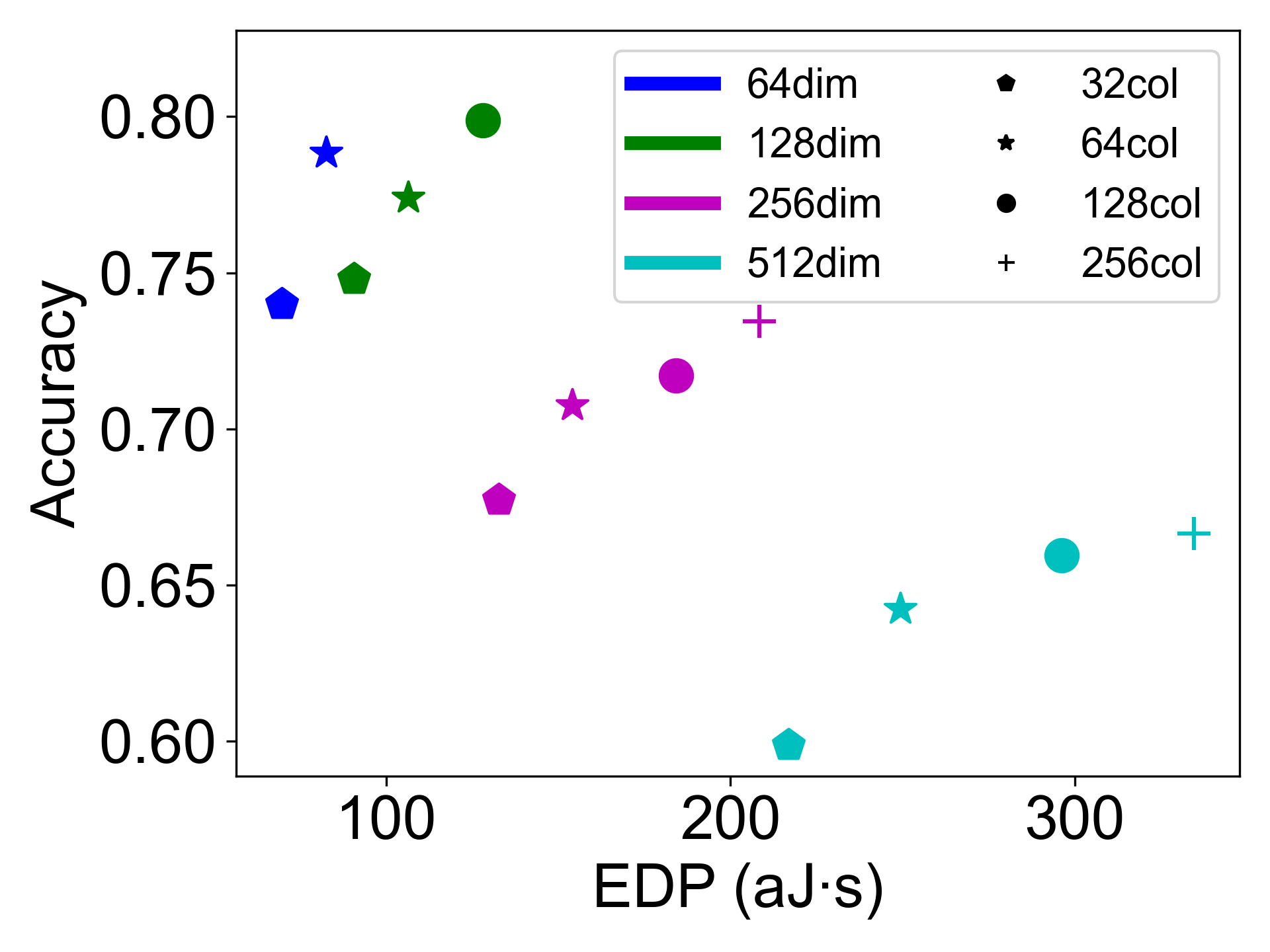}}
    \subfigure{\includegraphics[width=0.49\linewidth]{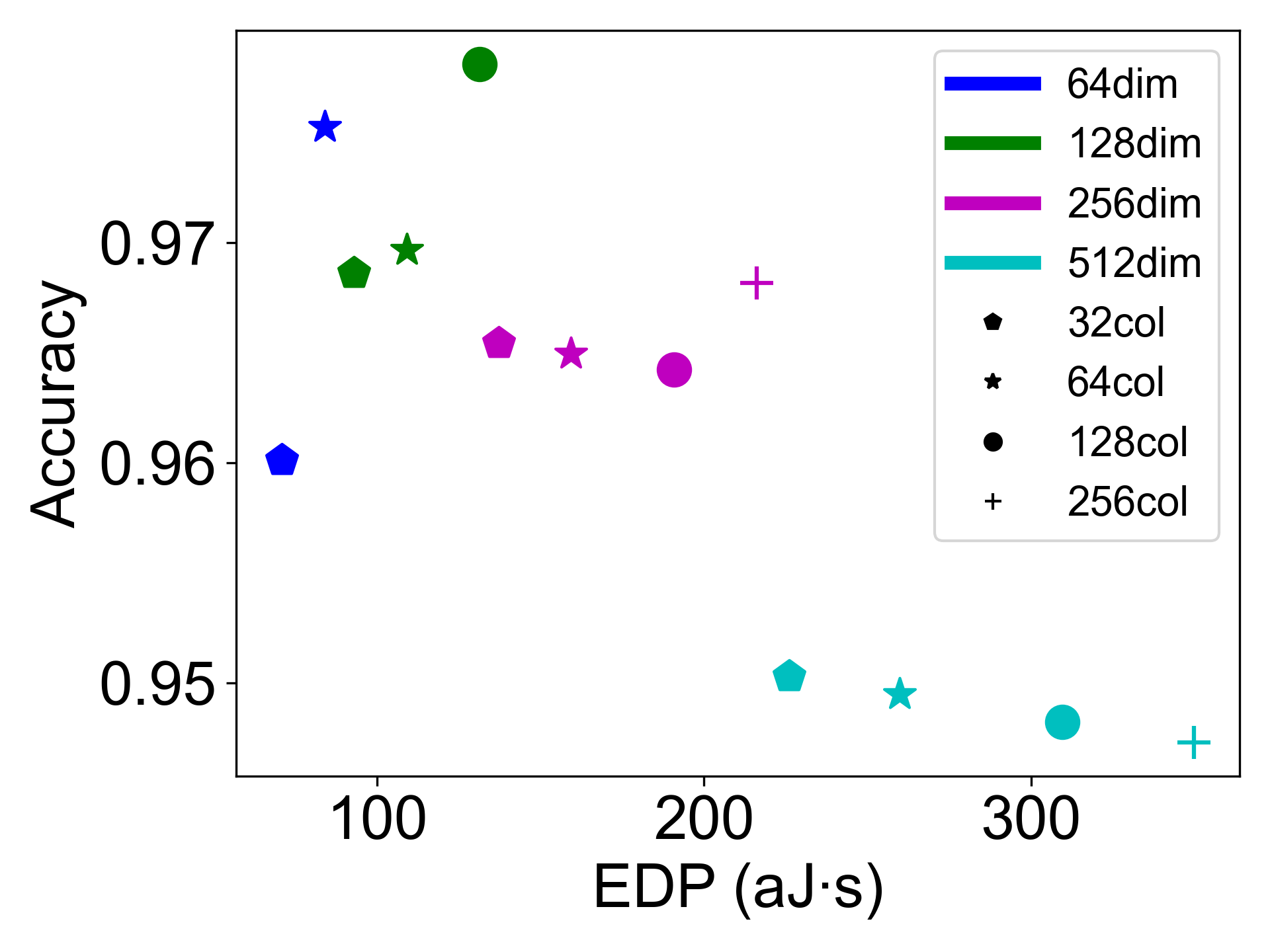}}
    \caption{Accuracy and EDP from \name with different number of embedding dimensions and CAM subarray column sizes. No quantization accuracy is 98.3\%. (Left) 2-bit quantization. (Right) 3-bit quantization.}
    \label{fig:acc-edp vs dim-col}
\end{figure}

\begin{figure}[t]
    \centering
    \vspace*{-5mm}
    \subfigure{\includegraphics[width=0.49\linewidth]{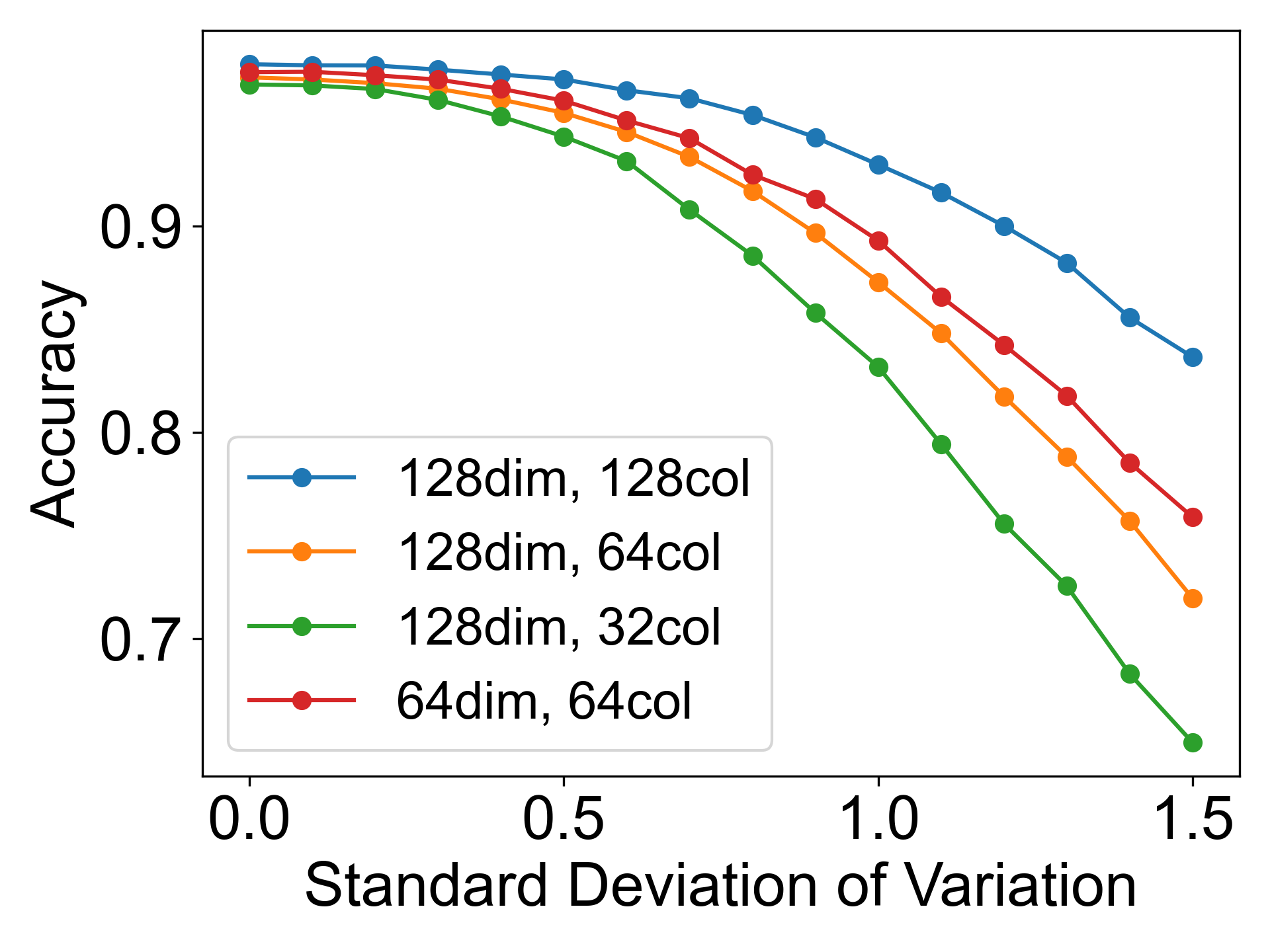}}
    \subfigure{\includegraphics[width=0.49\linewidth]{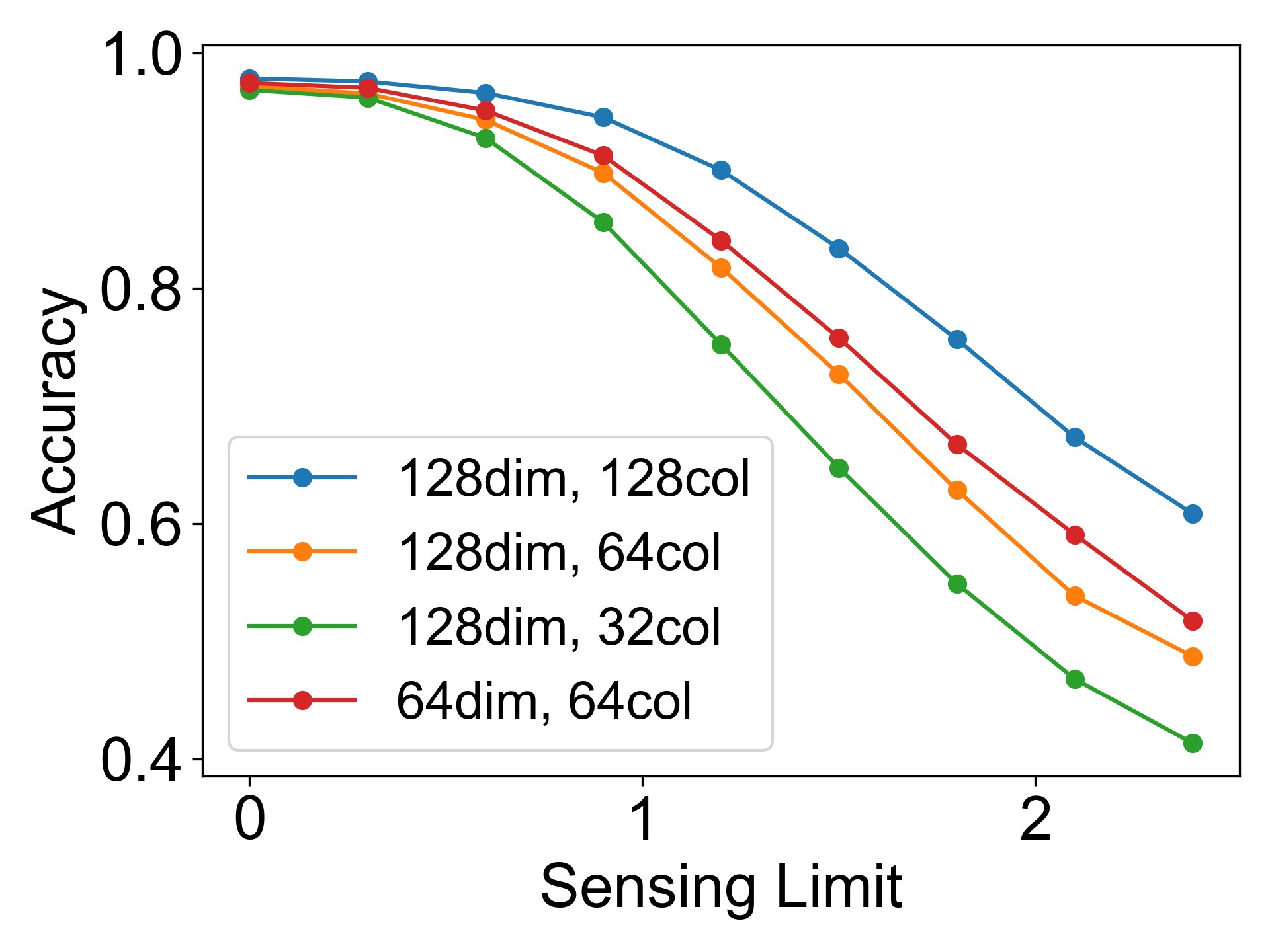}}
    \caption{Accuracy as a function of (left) device variation and (right) sensing limit for selected settings.}
    \label{fig:acc vs variation sensing limit}
\end{figure}


\subsection{Case Study}
In this subsection, we explore the design space using \name to study the MANN task in detail, from the application to the circuit level to reveal how these choices impact accuracy, latency, and energy. We adopt the same MANN model setup as in \cite{laguna2023fewshot}, which consists of a convolutional neural network (CNN) for generating embedding and the CAM-based accelerator for classification. In our study, we train the CNN models with different output embedding dimensions, ranging from 64 to 512, for 50000 epoches each.


\subsubsection{Impact of Data Type and Subarray Size}
We first evaluate the accuracy and EDP with different embedding dimensions, data types (i.e., quantization bits), and subarray sizes as shown in Fig.~\ref{fig:acc-edp vs dim-col}. Note that the number of embedding dimensions dictates whether horizontal voting-type merge is needed for a given subarray size (particularly number of columns). Fig.~\ref{fig:acc-edp vs dim-col} (left) shows that a 2-bit quantization hurts the accuracy by more than 18\%, while a 3-bit quantization (Fig. \ref{fig:acc-edp vs dim-col} (right)) only hurts the accuracy by less than 4\%. For the same subarray column size, we see that smaller dimensions generally perform better in terms of accuracy. Furthermore, for the same number of dimensions, larger subarrays also have higher accuracy. These trends are attributed to the increased error introduced by the utilized voting scheme. 

While EDP increases with a larger embedding dimension, a moderate 128-dimension embedding can reach the best accuracy for both quantization bits. For the same dimensions, EDP increases with larger subarrays due to increased latency (because of parasitics and less parallelism) since energy does not vary much. Also, we generally see that the accuracy and the EDP both increase with larger subarray size. Therefore, the comprehensive evaluation framework like \name could help designers to make deliberate tradeoff between accuracy and EDP.

\subsubsection{Impact of Circuit Non-idealities}
Here we modify the configuration in \name to evaluate the accuracy impacts of non-idealities including SL and D2D variation in a CAM-based system. Since a good design should have both high accuracy and low EDP, we choose 3-bit designs in Fig.~\ref{fig:acc-edp vs dim-col} whose accuracy are higher than 96.8\% (less than 1.5\% accuracy drop) and whose EDPs are less than 150aJ·s for this study. From the results shown in Fig. \ref{fig:acc vs variation sensing limit}, the two non-idealities incur similar impacts on accuracy. Specifically, for the same number of dimensions, a smaller subarray, even with better EDP, is less resilient to these non-idealities. While for the same number of columns, a smaller dimension is more vulnerable towards the non-idealities. This is mainly due to the fact the voting scheme, although effective under ideal conditions, gets worse faster when considering non-idealities.



\section{Conclusion}
\label{sec:conclusion}
In this work, we introduce \name, the first simulation framework for CAM-based in-memory search accelerators. It helps unravel the complexities associated with CAM-based accelerator design, thus offering a valuable tool for researchers and developers in the pursuit of efficient and accurate CAM-based solutions for current and future application domains. \name's modular design provides ample opportunities for future extension. 

\section*{Acknowledgment}
This project is supported in part by ACCESS – AI Chip Center for Emerging Smart Systems, sponsored by InnoHK funding, Hong Kong SAR, and by Semiconductor Research Corporation (SRC).

\bibliographystyle{./my_abbrv.bst}
\bibliography{references}

\end{document}